\documentclass{jfm}
\usepackage{graphicx}
\usepackage{epstopdf, epsfig}
\usepackage{gensymb}
\usepackage{subcaption}
\usepackage[dvipsnames]{xcolor}

\shorttitle{Evolution of the TNTI in stably stratified turbulent flows}
\shortauthor{M. Neamtu, D. Krug, J.P. Mollicone, M. van Reeuwijk, G. Haller and M. Holzner}

\title{Connecting the time evolution of the turbulence interface to coherent structures}

\author{Marius M. Neamtu-Halic\aff{1,2}
  \corresp{\email{neamtu@ifu.baug.ethz.ch}},
  Dominik Krug\aff{3},
	Jean-Paul Mollicone\aff{4},
	Maarten van Reeuwijk\aff{4},
  George Haller\aff{6}
 \and  Markus Holzner\aff{2,6}}

\affiliation{\aff{1}Institute of Environmental Engineering, ETH Zürich, CH-8039 Zürich, Switzerland
\aff{2}Swiss Federal Institute of Forest, Snow and Landscape Research WSL, 8903 Birmensdorf, Switzerland 
\aff{3}Physics of Fluids Group and Twente Max Planck Center, Department of Science and Technology, Mesa+ Institute, and J.M. Burgers Center for Fluid Dynamics, University of Twente, PO Box 217, 7500 AE Enschede, The Netherlands
\aff{4}Department of Civil and Environmental Engineering, Imperial College London, London SW7 2AZ, UK
\aff{5}Institute of Mechanical Systems, ETH Zürich, 8092 Zürich, Switzerland
\aff{6}Swiss Federal Institute of Aquatic Science and Technology Eawag, 8600 Dübendorf, Switzerland}
\begin{document}

\maketitle

\begin{abstract}

The surface area of turbulent/non-turbulent interfaces (TNTIs) is continuously produced and destroyed via stretching and curvature/propagation effects. Here, the mechanisms responsible for TNTI area growth and destruction are investigated in a turbulent flow with and without stable stratification through the time evolution equation of the TNTI area. We show that both terms have broad distributions and may locally contribute to either production or destruction. On average, however, the area growth is driven by stretching, which is approximately balanced by destruction by the curvature/propagation term. To investigate the contribution of different length scales to these processes, we apply spatial filtering to the data. In doing so, we find that the averages of the stretching and the curvature/propagation terms balance out across spatial scales of TNTI wrinkles and this scale-by-scale balance is consistent with an observed scale invariance of the nearby coherent vortices. Through a conditional analysis, we demonstrate that the TNTI area production (destruction) localizes at the front (lee) edge of the vortical structures in the interface proximity. Finally, we show that while basic mechanisms remain the same, increasing stratification reduces the rates at which TNTI surface area is produced as well as destroyed. We provide evidence that this reduction is largely connected to a change in the multiscale geometry of the interface, which tends to flatten in the wall-normal direction at all active length scales of the TNTI.

\end{abstract}


\section{Introduction}

In unbounded (e.g. jets, wakes, mixing layers) and semi-bounded (e.g. boundary layers) turbulent flows, a sharp and highly contorted interface separates the turbulent flow region from the non-turbulent ambient flow \citep{corrsin1955free,dimotakis2000mixing,da2014interfacial}. Across this so-called turbulent/non-turbulent interface (TNTI), surrounding irrotational fluid is continuously incorporated into the turbulent flow. This process, known as turbulent entrainment, is of importance in many practical applications, in that it governs the spreading rate, mixing and reactions in a wide range of industrial and environmental flows \citep{davidson2015turbulence, simpson1999gravity, murthy2013turbulent}. 

Commonly, the TNTI is identified through a threshold on a scalar quantity, such as vorticity magnitude or enstrophy \citep{bisset2002turbulent}, turbulent kinetic energy \citep{holzner2006generalized} or passive scalars \citep{westerweel2005mechanics}. From a local perspective, the entrained volume flux can be expressed as the product of ${\langle} v_{n} {\rangle}$, the average `local' entrainment velocity, where ${\langle} \cdot {\rangle}$ denotes an average over the surface area of the TNTI, and $A_{ \eta }$, the surface area of the TNTI \citep{sreenivasan1989mixing, mathew2002some}. To date, it is widely accepted that $A_{\eta}$ has a fractal shape \citep{sreenivasan1989mixing, de2013multiscale, krug2017fractal}, that bears the multiscale properties of turbulence, while its propagation velocity relative to the fluid elements ${\langle} v_{n} {\rangle}$  is very slow and on the order of the Kolmogorov velocity scale \citep{holzner2011laminar}. Although the local propagation of the TNTI is of viscous nature, it is well-known that the overall entrainment rate is independent of viscosity \citep{morton1956turbulent, townsend1966mechanism, tritton1988physical, tsinober2009informal}, viz. the Reynolds number. It thus follows that $A_{\eta}$ plays a crucial role in setting the entrainment rate, canceling out the viscosity dependency of ${\langle} v_{n} {\rangle}$ \citep{townsend1966mechanism}. To date, much of the research on the TNTI and associated entrainment process focused on vorticity transport across the TNTI \citep{westerweel2005mechanics, holzner2011laminar, silva2018scaling} and little is known about the mechanism that sets the surface area of the TNTI.

In his theoretical work, \citet{phillips1972entrainment} introduced an equation for the time evolution of the surface area of the TNTI (see \S\ref{sec:Methods}), which demonstrated that the growth of the interface area is the result of the sum between a flow stretching term and a curvature/propagation term. Hypothesizing a constant entrainment velocity over the TNTI, he concluded that on average the curvature/propagation effect creates TNTI area along the bulges and destroys it in the valleys of the TNTI. While this is an important theoretical finding, the local entrainment velocity is known to vary significantly along the TNTI \citep{holzner2011laminar, wolf2012investigations, watanabe2014enstrophy}, with a predominance of negative values implying entrainment that alternate with sporadic positive values representing detrainment zones \citep{wolf2012investigations, krug2017global, mistry2019kinematics}. In this work, we evaluate locally the stretching and the curvature/propagation terms on the TNTI in order to assess their role in the time evolution of surface area of the TNTI.

In the last decade, an effort has been made to define the role of coherent structures in the entrainment process. \citet{da2011role} used direct numerical simulation (DNS) data of a turbulent planar jet to show that the large-scale vortices near the TNTI define the shape of the interface area. Related findings were presented by \citet{lee2017signature}, who used conditional analysis to show that the surface area of the TNTI increases in the vicinity of large scales motions of a turbulent boundary layer. More recently, vortical structures near the TNTI have been shown to influence both the intensity of the local entrainment velocity and the mean curvature of the TNTI \citep{mistry2019kinematics, neamtu2019lagrangian}. This suggests that vortical structures may impact the evolution of the TNTI area. Besides, it was also observed that the coherent vortices near the TNTI distort the mean flow in their proximity \citep{lee2017signature, watanabe2017role}, which indicates that they may also influence the stretching of the TNTI. However, to date, the role of coherent flow structures on the time evolution of the TNTI area is largely unknown.
Contrary to previous approaches, it is our goal to identify Eulerian vortical structures in a systematic, observer-independent fashion. To this end we detect objective (i.e. frame-independent) Eulerian coherent structures (OECSs) \citep{haller2016defining,serra2016objective} and elucidate their role on the time evolution of the TNTI area.

Due to their relevance in many geophysical scenarios, turbulent flows with stable stratification have received substantial attention from the scientific community. Examples of such flows include river plumes \citep{macdonald2013heterogeneity}, cloud-top mixing layers \citep{mellado2010evaporatively} and oceanic overflows \citep{legg2009improving}. For these flows, the entrainment coefficient is known to diminish with increasing ratio between buoyancy and the shear strength of flow, represented by the Richardson number $Ri$ \citep{ellison1959turbulent}. Recently, it has been demonstrated \citep{ krug2015turbulent, van2018small} that the reduction of the entrainment coefficient with increasing $Ri$ is associated with the decrease of both ${\langle} v_{n} {\rangle}$ and $A_{\eta}$. In particular, \citet{krug2017fractal} showed that the reduction of $A_{\eta}$ is caused by the decrease of its fractal scaling exponent, while the scaling range remains largely unaffected. Here, we explore how varying $Ri$ affects the role of stretching and curvature/propagation for the time evolution of the TNTI area.

The main scope of the present work is to investigate the mechanisms that continuously produce and destroy the turbulence interface in flows with and without stable stratification with a particular regard to the role played in this process by coherent flow structures and the degree of stratification. 

The analysis will be carried out using a DNSs of a temporal wall-jet and gravity currents, details of which will be presented in \S\ref{sec:Methods}. This is followed by the presentation of the results in \S\ref{sec:results}, while concluding remarks are given in \S\ref{sec:discussion}.

\section{Methods}\label{sec:Methods}
\subsection{Direct numerical simulations}\label{subsec:DNS}

\begin{table}
  \begin{center}
\def~{\hphantom{0}}
  \begin{tabular}{ccccccc}
      $ $  & $\alpha (deg.)$ & $Ri_{0}$ & $Re_{0}$ & $Re_{\lambda}=\sqrt{15/ \nu \epsilon} e^{1/2}$ & $N_{x} N_{y} N_{z}$ & $L_{x} L_{y} L_{z} / h_{0}^{3}$\\[3pt]
       $Ri0$    & $-$ & $0$ & $3700$ & $115$ & $1536^{2} \times 1152$ & $20^{2} \times 10$\\[1ex]
       $Ri11$   & $10$ & $0.11$ & $3700$ & $105$ & $1536^{2} \times 1152$ & $20^{2} \times 10$\\[1ex]
       $Ri22$   & $5$ & $0.22$ & $3700$ & $70$ & $1536^{2} \times 1152$ & $20^{2} \times 10$\\[1ex]

  \end{tabular}
  \caption{Simulation parameters: $N_{i}$ and $L_{i}$ denote the number of grid points and the size along $i$-direction respectively.  The subscript $\textit{0}$ indicates the inflow parameters. Results for $Re_{\lambda}$ are averaged over $110<t<120$.}
\label{table:tab1}
  \end{center}
\end{table}

\begin{figure}
  \centerline{\includegraphics[width=1\linewidth]{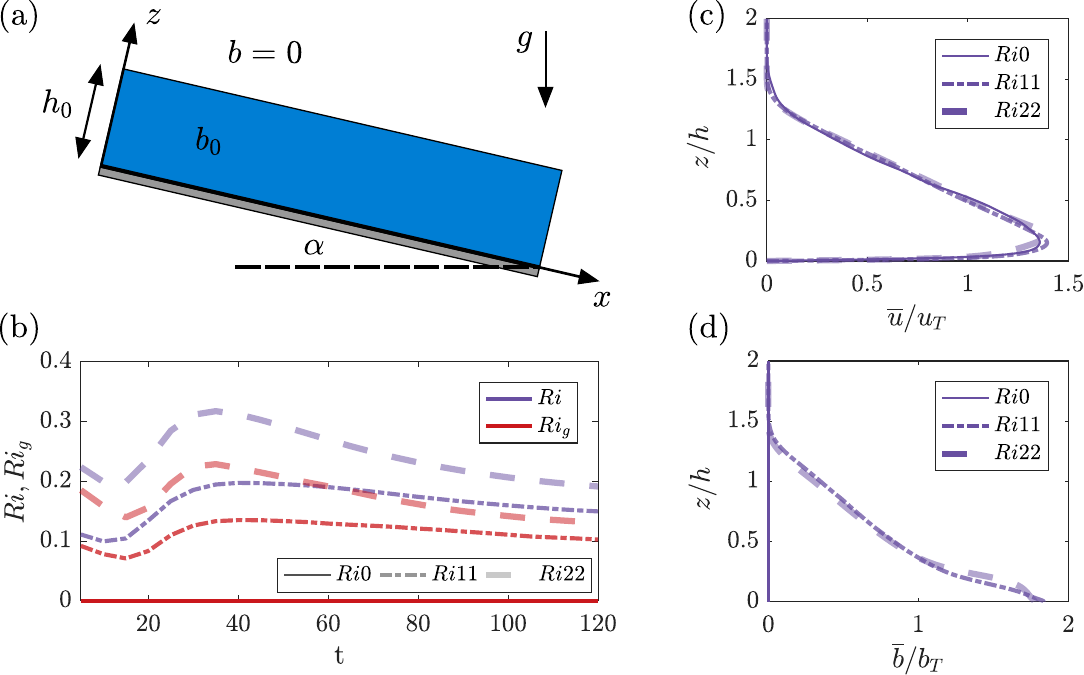}}
  \caption{Schematic representation of the simulation setup (a). Time variation of the bulk Richardson number (purple) and the gradient Richardson number (red) (b). Vertical profiles of the mean streamwise velocity (c) and mean buoyancy (d).}
\label{fig:fig1}
\end{figure}

In the present paper, we use DNSs of temporal gravity currents and of a temporal wall-jet for which there is no stratification. Temporally evolving flows are ideally suited for obtaining converged statistics relatively inexpensively since they are homogeneous in wall parallel planes \citep{van2018small}. For the simulations, we solve the Navier–Stokes equations in the Boussinesq approximation with a fourth-order accurate finite volume discretization scheme \citep{craske2015energy} on a cuboidal volume of $1536\times1536\times1152$ cells. Periodic boundary conditions are applied both in $y$ (the spanwise) and in $x$ (streamwise) directions. In the $z$ (vertical) direction, at the wall ($z=0$) and at the top of the simulation domain, no slip respectively free slip velocity boundary conditions are imposed for the velocity, whereas Neumann (no-flux) boundary conditions are imposed for buoyancy. As schematically represented in figure \ref{fig:fig1}(a), for the initial conditions (indicated by subscript 0), a uniform distribution of both the streamwise velocity $u_{0}$  and the buoyancy $b_{0}<0$ up to a height $h_{0}$ above the bottom wall is used.

The size of the domain in the $z$ direction is $L_{z}=10h_{0}$, whereas in $x$ and in $y$ directions it is $L_{x}=L_{y}=20h_{0}$.  
In order to simulate a sloping bottom, the buoyancy vector $ \textbf{\textit{b}}=b  \hat{ \textbf{\textit{g} } }$  is tilted at an angle $\alpha$ with respect to the vertical. Here $b$ is a scalar with Schmidt number $Sc=1$ and $\hat{ \textbf{\textit{g} } }=($sin $\alpha, 0, -$cos$ \alpha)g$. In this way, the component $b$ sin$(\alpha)$ drives the flow along $x$, while $b$ cos($\alpha$) is causing a stable stratification in the wall-normal direction. For a more detailed discussion on the DNSs concept and numerical configuration we refer to \citet{van2018small}, whereas the adequacy of the grid resolution is verified in  \citet{van2018mixing}. 

The different flow cases investigated here differ in the initial Richardson number $Ri_{0}=-B_{0}$cos($\alpha$)/$u_{0}^{2}$, whereas the initial bulk Reynolds number $Re_{0}=u_{0}h_{0}/\nu$, where $\nu$ is the kinematic viscosity, is kept constant.
 
To compute the time evolution of  $Ri=-B_{0}$cos($\alpha$)/$u_{T}^{2}$, we use the following top-hat definitions

\begin{equation}
  u_{T}h= \int_{0}^{\infty} \overline{u}dz, \,\,\,\,\,\,\,\,\,\,\,\,\,\,  u_{T}^{2} h= \int_{0}^{\infty} \overline{u}^{2}dz \,\,\,\,\,\,\,\,\,\,\,\,\,\, \textrm{and} \,\,\,\,\,\,\,\,\,\,\,\,\,\,  B_{0}=b_{T}h= \int_{0}^{\infty} \overline{b}dz,
\end{equation}
\\
where $B_{0}$ is a conserved quantity in the temporal problem \citep{van2018small} and $u$ is the streamwise velocity (the over-line indicates averaging in wall-parallel planes; the corresponding fluctuations are given by $u^{\prime}=u-\overline{u}$). The components of the velocity vector $\boldsymbol{u}$ along the $y$- and $z$-axes are denoted by $v$ and $w$ respectively. Table \ref{table:tab1} summarizes the parameters of the simulations employed in this study. To compute $Re_{\lambda}$, we average the turbulent kinetic energy $e$ and the rate of turbulent dissipation $\epsilon$ in horizontal planes, which were limited at $0.3<z/h<1.2$  in order to avoid the influence of the near-wall region \citep{krug2017fractal}. Note that the label of the flow cases indicates the value of $Ri_{0}$. For the gravity currents ($Ri11$ and $Ri22$), $Ri_{0}$ is varied by changing the inclination angle $\alpha$ while keeping the integral forcing sin($\alpha$)$ B_{0}$ in the $x$-direction constant. In addition, we ran a simulation with the buoyancy term switched off, resulting in an unstratified (temporal) wall jet ($Ri0$) that is driven by initial momentum only. Apart from the section \S\ref{subsec:TimeEvol}, where the whole domain was used, results will be based on data over six independent $xz$-planes, which are spaced equally in the $y$-direction, amounting to 250 snapshots over a period of 120$h_{0}/u_{0}$. Throughout the paper, the time $t$ is normalized by $h_{0}/u_{0}$.

For a flow characterization, the time evolution of the gradient Richardson number $Ri_{g}=(\overline{N}/\overline{S})^{2}$ is shown in figure \ref{fig:fig1}(b). In this definition, $\overline{N}^{2}=d\overline{b}/dz$ is the buoyancy frequency and $\overline{S}=d\overline{u}/dz$ is the mean shear. As can be seen from the figure, after an initial transient $Ri_{g}$ stabilizes and tends asymptotically towards two different constant values for the gravity current simulations. This behavior resembles that of the bulk $Ri$ however with slightly lower magnitudes (figure \ref{fig:fig1}b). It is important to note that for $Ri_{g}<1/4$, the flow is expected to be ‘shear-dominated’ according to the classification by \citet{mater2014unifying}, which has indeed been confirmed for the simulations presented here by \citet{krug2017fractal}. Moreover, as can be seen from figure \ref{fig:fig1}(c), when normalized with the top hat definitions, the mean streamwise velocity profiles of all the flow cases collapse on a single curve. This indicates that although there are fundamental differences between the stratified and unstratified cases, the structure of the flows is indeed very similar among all the flow cases. It is noteworthy that also the mean buoyancy profile of the gravity currents collapses on a single curve when normalized with the top hat definitions (figure \ref{fig:fig1}d).

\subsection{TNTI identification and local entrainment velocity}\label{subsec:TNTIvn}

In this study, the position of the TNTI is identified through a threshold on the enstrophy field $\omega^{2}=\omega_{i}\omega_{i}$, where $\omega_{i}$ is the vorticity vector \citep{bisset2002turbulent, holzner2007small, holzner2008lagrangian, silva2018scaling,wolf2012investigations, neamtu2019lagrangian}. Here, the threshold is the same for all flow cases and is set to $\omega_{thr}^2=10^{-3} {u_{0}}^2/{h_{0}}^2$, well within the interval of possible values identified by \citet{krug2017fractal} for DNSs with the same code and parameters as the ones presented here. By identifying the TNTI through an iso-surface of the enstrophy field, the entrainment velocity can be evaluated using the transport equation for the enstrophy \citep{holzner2011laminar, krug2015turbulent}. In this case $v_{n}$ is given by

\begin{equation}
  v_{n}  = -\frac{2 \omega_{i} \omega_{j} S_{ij}}{\vert \nabla \omega^{2} \vert} -\frac{2 \nu \omega_{i} \nabla^{2} \omega_{i}}{\vert \nabla \omega^{2} \vert} -\frac{2 \epsilon_{ijk} \omega_{i} \frac{\partial \textbf{\textit{b}}_{k}}{\partial x_{j}}}{\vert \nabla \omega^{2} \vert}, 
  \label{equation:eq3} 
\end{equation}
\\
where $\epsilon_{ijk}$ is the Levi-Civita operator and $S_{ij}=1/2(\partial u_{i}/\partial x_{j}+\partial u_{j}/\partial x_{i})$ is the strain rate tensor. 
Throughout the paper, we make use of both fully three-dimensional (3D) data, as well as two-dimensional (2D) data from vertical planes. In the case of the 2D approach, the entrainment velocity is determined through interface tracking, with a procedure similar to the one used by \citet{wolf2012investigations} in which, $v_{n}$ is computed as
\begin{equation}
  v_{n}=v_{I}-u_{n},
\end{equation}
\\
where $v_{I}$ is the local normal velocity of the TNTI and $u_{n}=\textbf{\textit{u}}_{f} \cdot \textbf{\textit{n}}$, with $\textbf{\textit{u}}_{f}$,  the flow velocity at the location of the TNTI and \textbf{\textit{n}}, the unit vector normal to the surface of the TNTI pointing towards the turbulent flow region. $v_{I}$ is computed by tracking the position of $\omega_{thr}^2$-isosurfaces in time.

\subsection{Equation for the time evolution of the TNTI area}\label{subsec:EqTimeEvol}
In the present work, we investigate in detail the time evolution of the TNTI area. Initially introduced by \citet{phillips1972entrainment}, the equation for the time evolution of a non-material infinitesimal surface element of area $\delta A$ reads  

\begin{equation}
  \frac{1}{\delta A} \frac{\textrm{d} \delta A}{\textrm{d}t} = (\delta_{ij}-n_{i}n_{j})S_{ij}+v_{n} \nabla \cdot \textbf{\textit{n}}, 
  \label{equation:eq2} 
\end{equation}
\\
where $\delta_{ij}$ is the Kronecker operator, $n_{i}$ are the components of $\textbf{\textit{n}}$, the unit vector normal to $\delta A$, $S_{ij}$ is the strain rate tensor and $\nabla \cdot \textbf{\textit{n}}$ is the mean curvature of the surface. In this study, $\textbf{\textit{n}}$ points outward towards the non-turbulent side. The first term on the right hand side (rhs) of (\ref{equation:eq2}) is the \emph{area stretch term}, whereas the second term, hereinafter referred to as \emph{curvature/propagation term}, arises from the combined effect of curvature and propagation velocity. Even though (\ref{equation:eq2}) was initially introduced in the context of studying TNTIs, it has since received more extensive attention from the reactive flows community \citep{candel1990flame,dopazo2006iso}. In this field, the equation (\ref{equation:eq2}) is known as the \emph{flame stretch equation}.

Since in this work, also 2D data from vertical planes is employed, we note that by passing from a 3D to a 2D approach, the TNTI reduces from a 2D surface to a 1D line and accordingly the symbol $A$ is substituted with the symbol $l$ for the length of a line element. In this case, the 1D analog of (\ref{equation:eq2}) reads

\begin{equation}
  \frac{1}{\delta l} \frac{\textrm{d} \delta l}{\textrm{d}t} = (\delta_{ij}-n_{i}n_{j})S_{ij}+v_{n} \nabla \cdot \textbf{\textit{n}}, 
  \label{equation:eq2d} 
\end{equation}
\\ 
 where $\textbf{\textit{n}}$ is the 2D unit vector normal to the segment $\delta l$ and $S_{ij}$ is the 2D strain rate tensor. 

\subsection{Coherent flow structures extraction}\label{subsec:OECSs}

For observer-independent vortical structure identification, we employ the  recently developed instantaneous vorticity deviation (IVD) technique of \citet{haller2016defining}. Derived  by \citet{haller2016dynamic} from a new, dynamic version of the classic polar decomposition, 
the IVD  field represents  an intrinsic material rotation rate in the fluid. Specifically, the IVD field, defined by the normed deviation of the vorticity vector $\boldsymbol{\omega}(\boldsymbol{x},t)$ from its spatial mean $\boldsymbol{\overline{\omega}} (t)$ over an evolving fluid mass, i.e., by the formula

\begin{equation}
  \textit{IVD}( \boldsymbol{x},t) =| \boldsymbol{\omega} ( \boldsymbol{x} ,t)- \boldsymbol{\overline{\omega}} (t)| 
  \label{equation:eqIVD} 
\end{equation}
\\ 
provides an observer-independent (objective) local angular velocity for each point of the  fluid mass. Outermost tubular surfaces of the IVD, therefore, enable the identification of OECSs in an observer-independent manner, as required for experimentally reproducible  coherent structure extraction \citep{haller2015lagrangian}. To find vortical OECSs in our data set,  we use the extraction algorithm developed in \citet{neamtu2019lagrangian}, applied here to the IVD field. In this case, the center of the vortical structure is represented by a codimension-2 line that is the concatenation of local maxima of the IVD in planes perpendicular to the line itself, whereas the boundary of the structure is the union of the outermost almost-convex iso-contours of IVD encircling the local maxima in planes perpendicular to the center-line.

When 2D data from vertical planes is considered, a 2D OECS results form the intersection of a 3D structure with the plane itself. In this case we select only those OECSs with a limited intersection angle with respect to the normal unit vector of the plane. To this end, we impose an upper limit to the ratio between the two eigenvalues of the Hessian of IVD at the location of IVD maxima. The rationale behind this selection is based on the fact that most of the dynamics of tubular vortical structures happens in planes perpendicular to the center-line of the structure.

\section{Results}\label{sec:results}
\subsection{Time evolution of the TNTI area}\label{subsec:TimeEvol}

\begin{figure}
  \centerline{\includegraphics[width=1\linewidth]{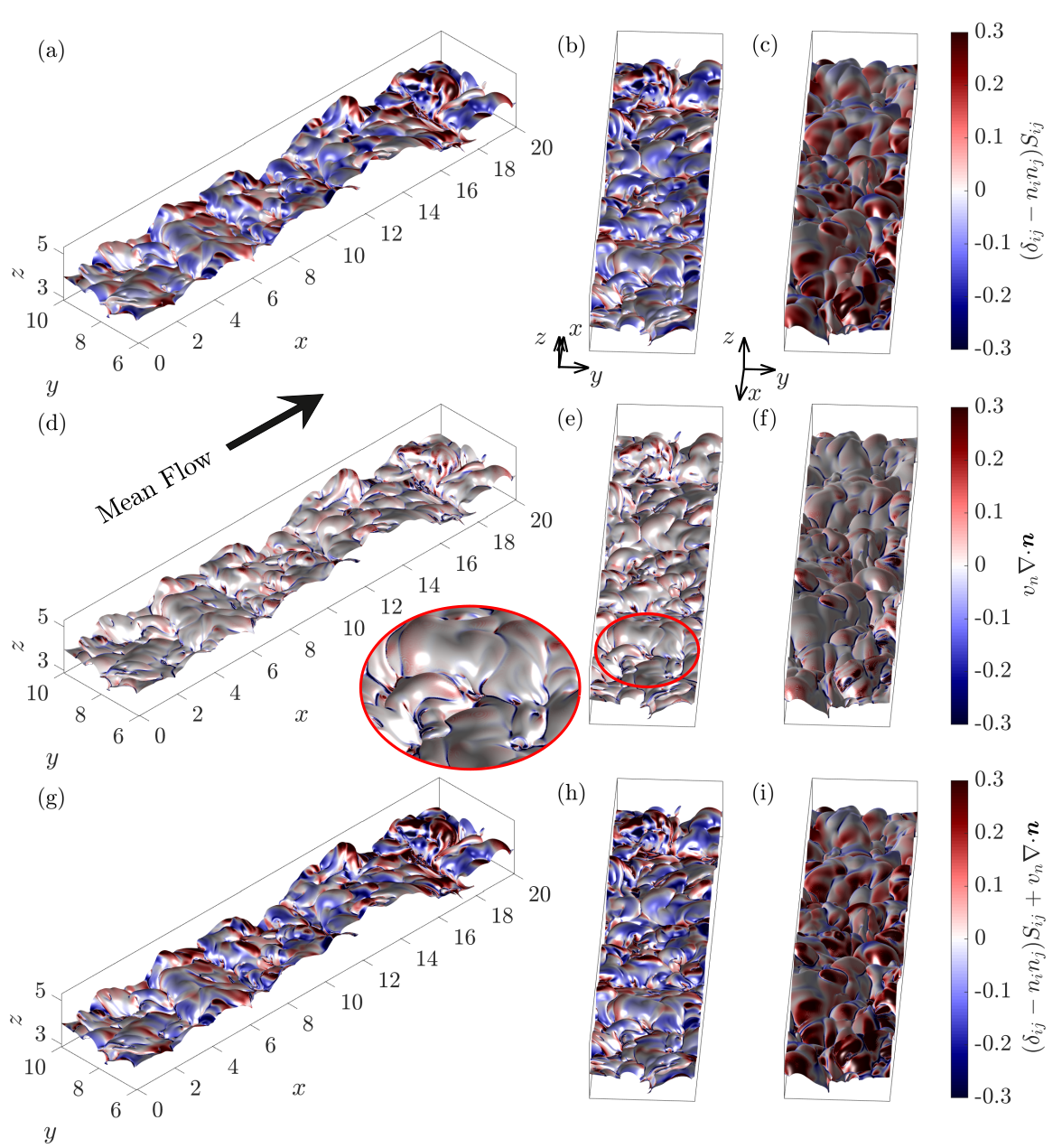}}
  \caption{Visualization of the stretching term (a-c), curvature/propagation term (d-f) and the time evolution of a non-material infinitesimal element of area $\delta A$,  $1/ \delta A \cdot \textrm{d}(\delta A)/\textrm{d}t$ (g-h) of the TNTI for different view angles as captured from a snapshot of the $Ri11$ flow case at $t=100$. The black arrow indicates the mean flow direction.}
\label{fig:fig2}
\end{figure}

A visualization of the TNTI in a sub-domain of the gravity current for the $Ri11$ flow case at $t=100$ is shown in figure \ref{fig:fig2}. Here, the TNTI is color-coded with the terms of (\ref{equation:eq2}). In equation (\ref{equation:eq2}), positive values of the terms on the right-hand side contribute to the production of surface area of the TNTI, whereas negative values promote its destruction. As can be seen from figure \ref{fig:fig2}(a-f), both $(\delta_{ij}-n_{i}n_{j})S_{ij}$ and $v_{n} \nabla \cdot \textbf{\textit{n}}$ act to produce and destroy the TNTI area. In particular, $(\delta_{ij}-n_{i}n_{j})S_{ij}$ is mainly positive at the leading edges (figure \ref{fig:fig2}b) and negative at the trailing edges of the bulges of the TNTI  (figure \ref{fig:fig2}c). 

Conversely, $v_{n} \nabla \cdot \textbf{\textit{n}}$  is particularly active in the valleys of the interface (see e.g. figure \ref{fig:fig2}(e) and \ref{fig:fig2}(f) and the zoom of figure \ref{fig:fig2}(e)), where strong negative values can be observed. This is expected, given that the valleys of the TNTI are regions with high curvature. Moreover, $v_{n} \nabla \cdot \textbf{\textit{n}}$ appears to have positive values on the bulges, but at a lower intensity as compared to the valleys. The sum of the two terms, which describes the time evolution of a non-material infinitesimal element of area $\delta A$, is mostly dominated by $(\delta_{ij}-n_{i}n_{j})S_{ij}$ on the bulges and by $v_{n} \nabla \cdot \textbf{\textit{n}}$  in the valleys (see figure \ref{fig:fig2}h  and i). Interestingly, we note that the sign of the patches, especially for $(\delta_{ij}-n_{i}n_{j})S_{ij}$  and $1/ \delta A \cdot \textrm{d}(\delta A) / \textrm{d}t$, seems to correlate with the geometry of the bulges. 

\begin{figure}
  \centerline{\includegraphics[width=0.85\linewidth]{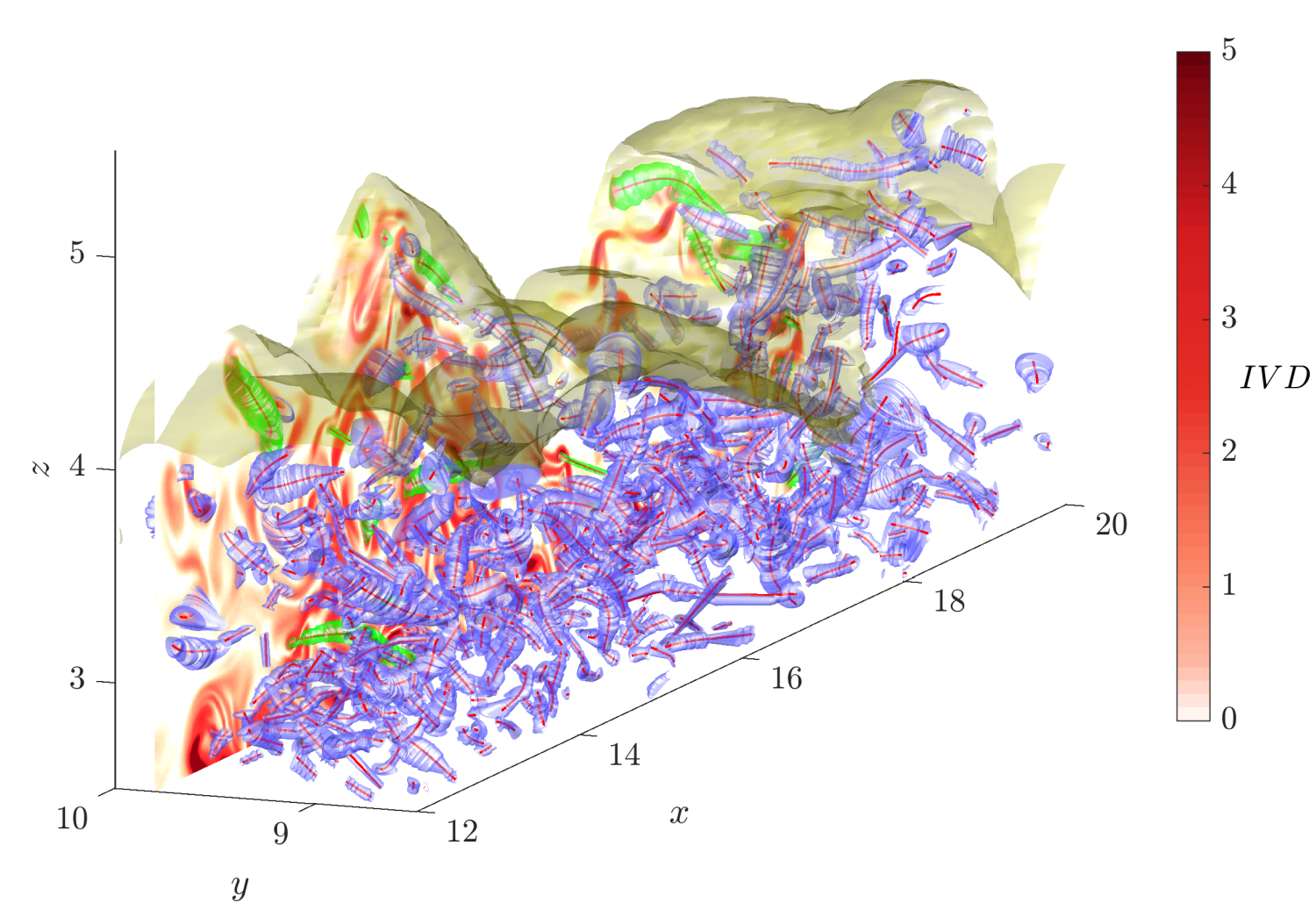}}
  \caption{Visualization of the OECSs and the TNTI from $Ri11$ at $t=100$. The OECSs are represented by blue tubular surfaces (boundaries) surrounding one-dimensional curves (centers), whereas the TNTI is represented by the yellow open-surface. The region above the TNTI is irrotational, whereas below the surface, the flow is turbulent. The OECSs with green boundaries cross the vertical plane at $y=9.75$ almost perpendicularly. On the vertical plane at $y=9.75$ the IVD field is shown in red contourplot.}
\label{fig:fig3}
\end{figure}

From figure \ref{fig:fig2}, there appears to be a spatial organization of the terms in equation (\ref{equation:eq2}) with respect to the TNTI bulges. \citet{neamtu2019lagrangian} showed that TNTI bulges are populated by OECSs. In figure \ref{fig:fig3}, we show part of the OECSs extracted from a subvolume of the flow field shown in figure \ref{fig:fig2}.

Here, the OECSs are composed by tubular surfaces, that constitute the boundaries of the structures. These surfaces enclose 1D-curves, which represent the center of the structures. In addition to the OECSs, we also display the nearby TNTI (yellow transparent open-surface) along with a vertical $xz$-plane (at $y=9.75$) color-coded with intensity of the IVD. As observed in \citet{neamtu2019lagrangian}, most of the bulges are filled with OECSs that appear to shape the nearby TNTI. To investigate the local effect of the OECSs on the TNTI area production/destruction, we use the conditional analysis of \citet{neamtu2019lagrangian}, and explore the impact of the coherent structures on $(\delta_{ij}-n_{i}n_{j})S_{ij}$ and $v_{n} \nabla \cdot \textbf{\textit{n}}$.
 
To simplify the analysis and manage the computational cost, we perform the subsequent analysis in 2D. Using the selection criterion described in \S\ref{subsec:OECSs}, in figure \ref{fig:fig3}, we highlight the 3D OECSs (green boundaries) that are considered for the further 2D analysis in the case of the vertical plane at $y=9.75$.

\begin{figure}
  \centerline{\includegraphics[width=0.9\linewidth]{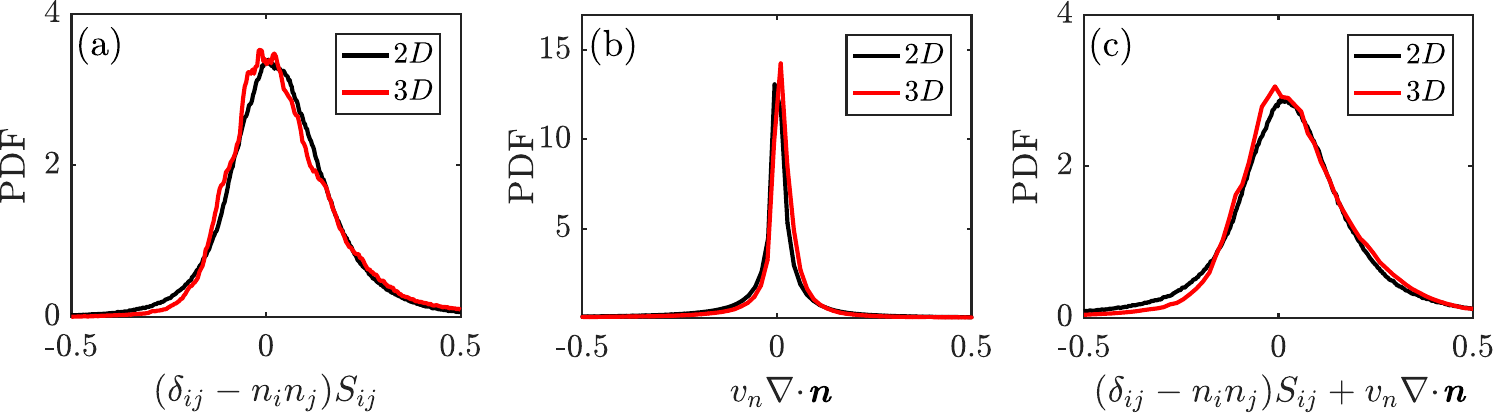}}
  \caption{PDFs of stretching (a) and curvature/propagation terms (b) and their sum (c) from 2D (black) and 3D data (red) for $Ri11$.}
\label{fig:fig4}
\end{figure}

Before proceeding, the accuracy of the 2D approach as compared to the 3D approach is addressed in terms of probability density functions (PDFs) of the three terms of (\ref{equation:eq2}). In figure \ref{fig:fig4}, we show the results for the $Ri11$ flow case. As can be seen from the figure, in all three PDFs, the two approaches provide very similar results. 

In general, the PDF of the stretching term (figure \ref{fig:fig4}a) has a higher positive tail and the overall distribution is slightly shifted towards positive values. Conversely, the PDF of the curvature/propagation term (figure \ref{fig:fig4}b) has a higher negative tail and has a strong peak at $v_{n} \nabla \cdot \textbf{\textit{n}}=0$. Moreover, the variance of the two aforementioned PDFs is different. In fact, the PDF of the curvature/propagation term presents a much narrower distribution as compared to the PDF of the stretching term. The PDF of the sum of the two terms (figure \ref{fig:fig4}c) shows characteristics of the PDFs of both $(\delta_{ij}-n_{i}n_{j})S_{ij}$ and $v_{n} \nabla \cdot \textbf{\textit{n}}$, in that it has a positive peak and a slightly higher negative tail.

\subsection{Time evolution of the TNTI area: a 2D approach}\label{subsec:2DTimeEvol}

\begin{figure}
  \centerline{\includegraphics[width=0.9\linewidth]{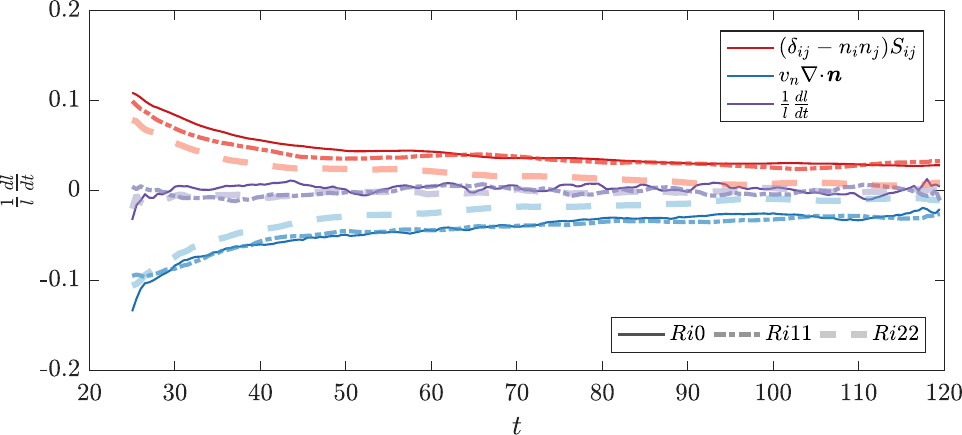}}
  \caption{Time evolution of the spatial average of stretching term (red), curvature/propagation term (blue) and $1/\delta l \cdot \textrm{d}(\delta l)/\textrm{d}t$ (purple) for $Ri0$ (continuous line), $Ri11$ (dash-dotted line) and $Ri22$ (dashed line).}
\label{fig:fig5}
\end{figure}

In the following, only results from 2D data are presented. 
In figure \ref{fig:fig5}, we show the time evolution of the spatial averages of the terms in (\ref{equation:eq2}) for each of the flow cases. Note that, $1/\delta l \cdot \textrm{d}(\delta l)/\textrm{d}t$ can be computed locally as a sum of $(\delta_{ij}-n_{i}n_{j})S_{ij}$ and $v_{n} \nabla \cdot \textbf{\textit{n}}$ only. However, since we consider average values over the entire box, an estimation of the average of $1/\delta l \cdot \textrm{d}(\delta l)/\textrm{d}t$ can be made by taking for $\delta l$ the entire length of TNTI. This can be formalized according to

\begin{equation}
  {\Big \langle}\frac{1}{\delta l} \frac{\textrm{d} \delta l}{\textrm{d}t}{\Big \rangle} =\frac{1}{\sum_{l} \delta l} \frac{\textrm{d} (\sum_{l} \delta l)}{\textrm{d}t}= {\langle}(\delta_{ij}-n_{i}n_{j})S_{ij}{\rangle}+{\langle}v_{n} \nabla \cdot \textbf{\textit{n}}{\rangle}. 
  \label{equation:eq2d_av} 
\end{equation}
\\
As can be seen for all the flow cases, the stretching term is positive on average and decays in time from about $0.1$ at $t=25$ to $0.01$ at $t=120$. Conversely, the curvature/propagation term is negative on average and its magnitude decays in time similar to the stretching term from about $-0.1$ at $t=25$ to $-0.01$ at $t=120$. We note that several turbulent time scales were tested for the scaling of these trends, namely, the Kolmogorov time scale $(\nu/\epsilon)^{1/2}$, the mean shear time scale $\overline{S}^{-1}$, the turbulence time scale $e/\epsilon$ and the integral time scale $h/u_{T}$. However, none of these time scales were able to collapse rates in time and across $Ri$ hinting thus at a multiscale nature of the terms in (\ref{equation:eq2}). As $Ri$ increases, both the average stretching and curvature/propagation terms decrease. Moreover, the spatial average growth of the TNTI surface, $1/\delta l \cdot \textrm{d}(\delta l)/\textrm{d}t$, results to be very small for all the time steps between $t=25$ and $t=120$. That is, the two terms on the rhs of equation (\ref{equation:eq2}) approximately balance each other out for all the time steps shown in figure \ref{fig:fig5}.

\begin{figure}
  \centerline{\includegraphics[width=1\linewidth]{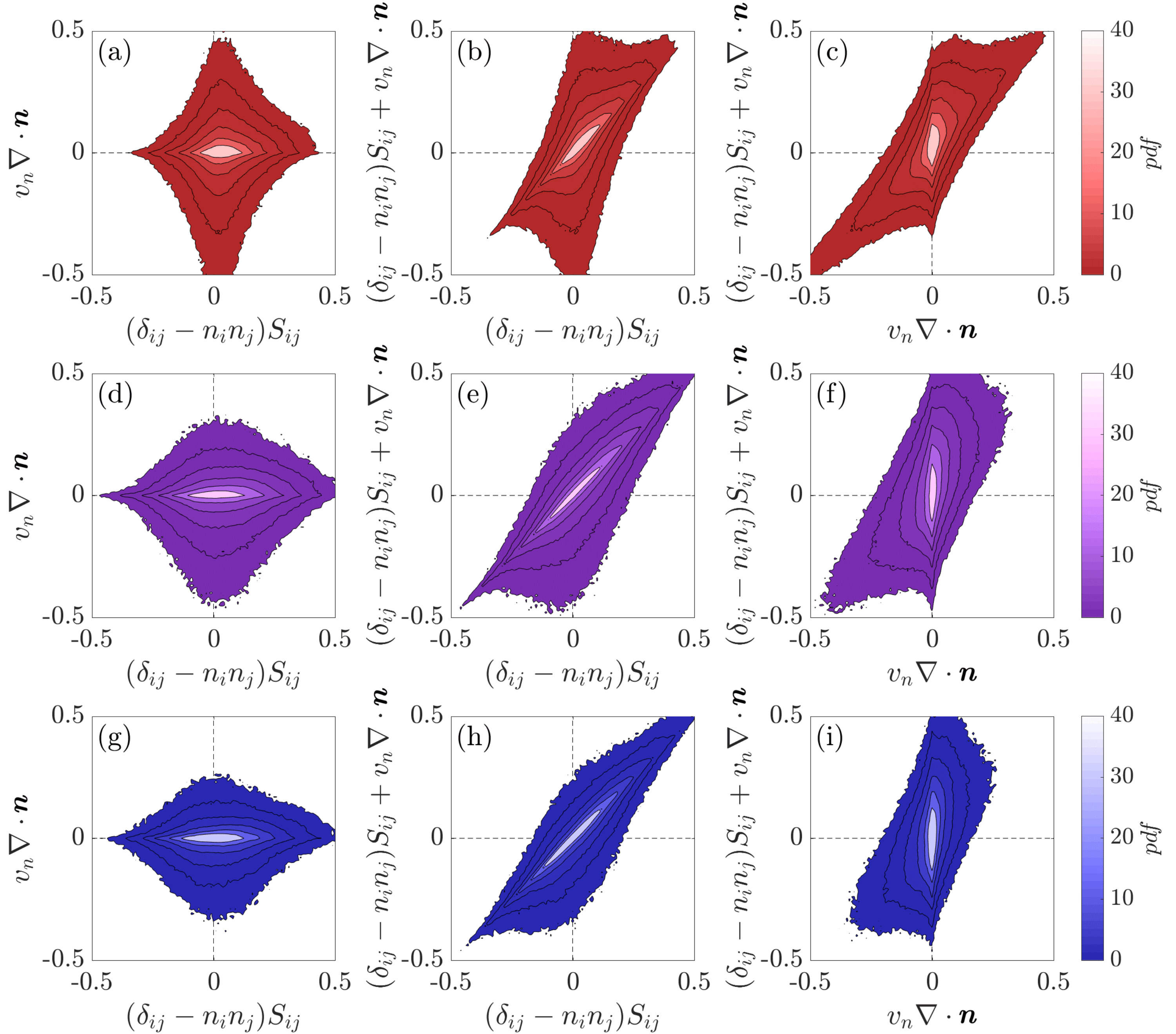}}
  \caption{Joint PDFs of $(\delta_{ij}-n_{i}n_{j})S_{ij}$, $v_{n} \nabla \cdot \textbf{\textit{n}}$ and their sum for $Ri0$ (a-c), $Ri11$ (d-f) and $Ri22$ (g-h) for $30<t<120$. The corresponding values of the black contour lines increase with logarithmic intervals.}
\label{fig:fig6}
\end{figure}

In order to understand how the two terms on the rhs of equation (\ref{equation:eq2}) balance out on average, we show in figure \ref{fig:fig6} the joint PDFs of all possible couples of the terms in equation (\ref{equation:eq2}) for $Ri0$ (a-c), $Ri11$ (d-f) and $Ri22$ (g-h). 

The joint PDF of $(\delta_{ij}-n_{i}n_{j})S_{ij}$  and  $v_{n} \nabla \cdot \textbf{\textit{n}}$ (first column of figure \ref{fig:fig6}) has a vertically elongated shape with a distinguishable horizontally elongated peak. This demonstrates that the stretching term dominates over the curvature/propagation term for small values of $(\delta_{ij}-n_{i}n_{j})S_{ij}$  and  $v_{n} \nabla \cdot \textbf{\textit{n}}$ (between $\pm 0.1$), whereas the curvature/propagation term has longer tails. Furthermore, the joint PDF shows higher probability for positive values of $(\delta_{ij}-n_{i}n_{j})S_{ij}$  and negative values of $v_{n} \nabla \cdot \textbf{\textit{n}}$. As $Ri$ increases the tails of the curvature/propagation term are reduced, whereas the stretching term shows a slightly broader distribution and a more centered peak. The joint PDF of $(\delta_{ij}-n_{i}n_{j})S_{ij}$ and $1/\delta l \cdot \textrm{d}(\delta l)/\textrm{d}t$ (sum of the two terms on rhs of (\ref{equation:eq2})) is shown in the second column of figure \ref{fig:fig6}.  A high degree of correlation between the $(\delta_{ij}-n_{i}n_{j})S_{ij}$ and $1/\delta l \cdot \textrm{d}(\delta l)/\textrm{d}t$ can be noticed for small values (between $\pm 0.1$), whereas for higher values the PDF is elongated in the vertical direction, a sign of weaker correlation between the two terms. Again, as $R$i increases the tails of $1/\delta l \cdot \textrm{d}(\delta l)/\textrm{d}t$ reduce. The joint PDF between $v_{n} \nabla \cdot \textbf{\textit{n}}$  and $1/\delta l \cdot \textrm{d}(\delta l)/\textrm{d}t$ is shown in the last column of figure \ref{fig:fig6}. In this case, the two quantities correlate very well for intense values, whereas they appear to be almost uncorrelated near origin. In conclusion, the area growth is mostly driven by the strain term for weak to moderate events that tends to produce interface area. However, the large tails of the curvature/propagation term dominate the extreme events and has a negative mean, so that on average this term counterbalances the positive stretching. As $Ri$ increases the curvature/propagation term has a narrower distribution, whereas the peak of the stretching term tends to move closer to the origin. A physical interpretation of this latter observation is provided in \S\ref{subsec:OECSsTNTI}, where we connect the interface evolution to the presence of coherent structures.

\subsubsection{Effect of the stable stratification on the production/destruction process of the TNTI area}\label{subsubsec:StableStrat}

In the following we investigate the effect of the stable stratification on the terms of (\ref{equation:eq2}). Initially, we focus on the stretching term, that in 2D can be written as:
\begin{equation}
 (\delta_{ij}-n_{i}n_{j})S_{ij}=(1-n_{x}^2) \frac{\partial u}{\partial x}-n_{x}n_{z}(\frac{\partial u}{\partial z}+\frac{\partial w}{\partial x}) +(1-n_{z}^2) \frac{\partial w}{\partial z}. 
  \label{equation:eq3} 
\end{equation}

\begin{figure}
  \centerline{\includegraphics[width=0.95\linewidth]{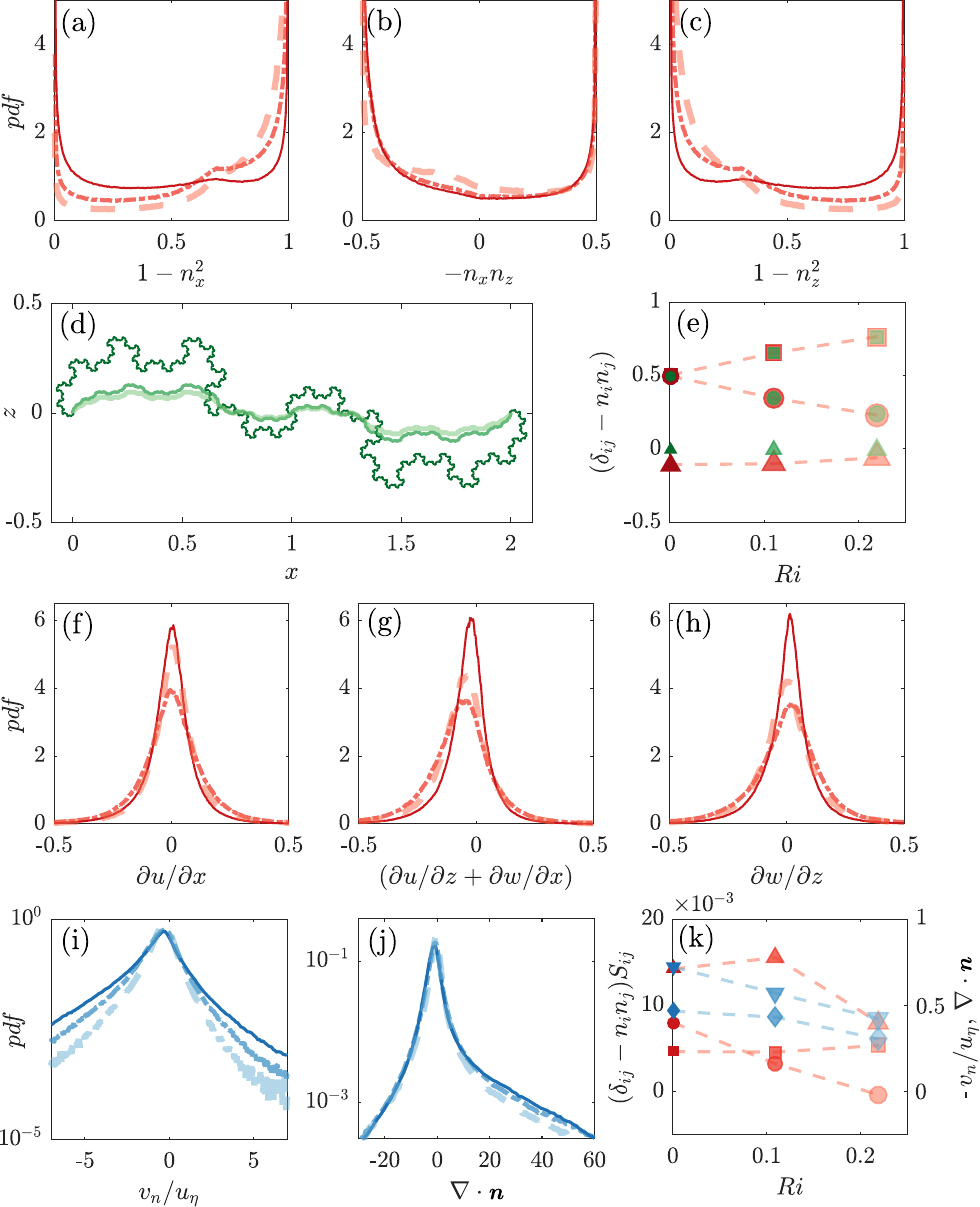}}
  \caption{PDFs of the three components of $(\delta_{ij}-n_{i}n_{j})$ (a-c) for $Ri0$ (continuous line), $Ri11$ (dash-dotted line) and $Ri22$ (dashed line). TNTI surface as obtained from the fractal model of \citep{krug2017fractal} for $r=0.3$ ($Ri0$), $r=0.12$ ($Ri11$) and $r=0.09$ ($Ri22$) (d). The thickness of the lines increases with $Ri$. Average value of $(1-n_{x}^2)$ (squares), $-n_{x}n_{z}$ (triangles) and  $(1-n_{x}^2)$ (circles) from the fractal model (green) and form the DNS data (red) with $30<t<120$ (e).  PDFs of the three rate of strain components (f-h) of the stretching term and PDFs of the entrainment velocity (i) and of the curvature of the TNTI (j) for $30<t<120$.  Average values of  $(1-n_{x}^2)\partial u/\partial x$ (square symbols), $-n_{x}n_{z}(\partial u/\partial z+\partial w/\partial x)$ (red triangle symbols), $(1-n_{z}^2)\partial w/\partial z$ (circles) and of $\nabla \cdot \textbf{\textit{n}}$ (blue diamonds) and $-v_{n}/u_{\eta}$  (blue triangles) against the $Ri$ number (j).}
\label{fig:fig7}
\end{figure}

In figure \ref{fig:fig7}, we show PDFs of the three components of $(\delta_{ij}-n_{i}n_{j})$ and of the three components of $S_{ij}$. As can be seen, a significant effect of the stable stratification can be noticed on the three coefficients $(\delta_{ij}-n_{i}n_{j})$ of (\ref{equation:eq3}) (figure \ref{fig:fig7}a-c). While $(1-n_{x}^2)$ shows a higher probability for values close to unity with increasing $Ri$ (figure \ref{fig:fig7}a), $(1-n_{z}^2)$ is seen to diminish as the stratification increases (figure \ref{fig:fig7}c). Also, as $Ri$ increases, $-n_{x}n_{z}$ shows a higher probability for values close to 0. These observations indicate that $n_{x} \rightarrow 0$ while $n_{z}\rightarrow1$ as the stratification increases, that is, the interface tends to flatten with increasing $Ri$. 

To better understand the effect of the stratification on the components of $(\delta_{ij}-n_{i}n_{j})$ tensor, we use the fractal scaling theory for the geometry of the TNTI \citep{sreenivasan1989mixing}. According to the theory, the length of the TNTI depends on $l_{i}/l_{o}$, the inner and the outer cutoffs of the scaling range and on $\beta$, the fractal scaling exponent. In their work, \citet{krug2017fractal} showed that for gravity currents, $l_{i}/l_{o}$ is essentially constant for $0<Ri<0.22$, whereas $\beta$ decreases with increasing stratification. Moreover, the authors observed that the convolutions of the TNTI are anisotropic, scaling with $l_{sk}$ in the wall-normal and with $h$ in the streamwise direction. Based on the observation that the ratio $r=l_{sk}/h$ decreases with increasing stratification, they implemented a model for $\beta=f(r)$, that was able to reproduce the trends of the fractal scaling exponent $\beta$ with increasing $Ri$. The model is based on a simple fractal model where in subsequent iterations line segments with length $l_{n+1}$ are placed at distance $rl_{n}$ from the center of $l_{n}$ \citep{krug2017fractal}. Here we use this model to check quantitatively if the trends observed in figure \ref{fig:fig7}(a-c) can also be related to the anisotropy of the interface bulges. In figure \ref{fig:fig7}(d), we display the geometry of the modeled interface where $l_{i}/l{o}\approx 100$, chosen according to our Reynolds number, and $r$ has the values indicated in the caption. As can be noticed, for $Ri22$ the fluctuations of the TNTI position in the wall-normal direction are much lower as compared to the wall-jet and thus the interface tends to flatten with increasing stratification. A comparison between the fractal model and the DNS data in terms of average value of $(\delta_{ij}-n_{i}n_{j})$ components against $Ri$ is shown in figure \ref{fig:fig7}(e). Notably, the model reproduces rather well the average of $(\delta_{ij}-n_{i}n_{j})$, especially for $(1-n_{x}^2)$ and $(1-n_{z}^2)$, the two terms that vary more significantly with $Ri$. We therefore conclude that, in agreement with the model, the stratification impacts the interface geometry at all `active' length scales of the TNTI between $l_{i}$ and $l_{o}$.

In figure \ref{fig:fig7}(f-h), we show PDFs of the components of the rate of strain. All three PDFs are only weakly affected by increasing stratification, which is reflected in a moderate increase of the weight of the tail and an associated decrease of the peak at small magnitudes. While in the PDFs of $\partial u/\partial x$ and $\partial w/\partial z$ the skewness reduces (figure \ref{fig:fig7}f and h), the PDF of $\partial u/\partial z+ \partial w/\partial x$ is more negatively skewed with increasing $Ri$. The latter is consistent with a stronger mean velocity gradient, i.e. smaller $h$ at similar $u_{T}$, and therefore stronger $\partial u/\partial z$ at increasing $Ri$.

The average values of the terms in equation (\ref{equation:eq3}) against $Ri$ are shown in figure \ref{fig:fig7}(k). Although the PDF of $\partial u/\partial x$ presents a slightly higher negative tail, the average of $(1-n_{x}^2)\partial u/\partial x$ is positive. This means that there is coupling between high values of $(1-n_{x}^2)$ and positive values of $\partial u/\partial x$. As the stratification increases, $(1-n_{x}^2)\partial u/\partial x$ increases slightly. The average of $-n_{x}n_{z}(\partial u/\partial z+\partial w/\partial x)$ is positive as expected from the PDFs in figure \ref{fig:fig7}(b) and (c). As the stratification increases, $-n_{x}n_{z}(\partial u/\partial z+\partial w/\partial x)$ increases slightly at $Ri=0.11$, to decrease afterwards at $Ri=0.22$. In particular, the smaller value $-n_{x}n_{z}(\partial u/\partial z+\partial w/\partial x)$ at $Ri=0.22$ might be related to a higher probability of $-n_{x}n_{z}=0$ as compared to the other flow cases, given that the PDF of $\partial u/\partial z+\partial w/\partial x$ for $Ri22$ is comparable to the one of $Ri11$ and exhibits even a higher negative tail as compared to the one of $Ri0$. Also the average of $(1-n_{z}^2)\partial w/\partial z$ is positive, meaning that, as for the other terms, there is a coupling of high values of $1-n_{z}^2$ and positive values of $\partial w/\partial z$. As the $Ri$ increases, $1-n_{z}^2$ tends towards smaller values and the average of $(1-n_{z}^2)\partial w/\partial z$ decreases.

In conclusion, most of the reduction in the stretching term of (\ref{equation:eq2}) with increasing stratification is related to a change in the components of $(\delta_{ij}-n_{i}n_{j})$ tensor as a result of a change in the multiscale geometry of the TNTI which tends to flatten out with increasing $Ri$.

Furthermore, in figure \ref{fig:fig7} we show the impact of the stable stratification on the curvature/propagation term. As can be gleaned from the presented PDFs, the stable stratification reduces the magnitude of both $v_{n}$ and $\nabla \cdot \textbf{\textit{n}}$. In particular, while the reduction of the $v_{n}$ (figure \ref{fig:fig7}i) with increasing $Ri$ is well-documented \citep[see e.g.][]{krug2015turbulent, van2018mixing}, it can be seen from figure \ref{fig:fig7}(j) that the stratification reduces also the probability to observe strong convex regions, associated with positive values of $\nabla \cdot \textbf{\textit{n}}$. This is not unexpected, given the changes in the geometry of the TNTI discussed above. The average values of the two terms against $Ri$ is shown in figure \ref{fig:fig7}(k) and as expected the magnitude of both the terms decrease on average with increasing stratification.

\subsubsection{Multiscale nature of the production/destruction of the TNTI area}\label{subsubsec:Multiscale}

\begin{figure}
  \centerline{\includegraphics[width=0.8\linewidth]{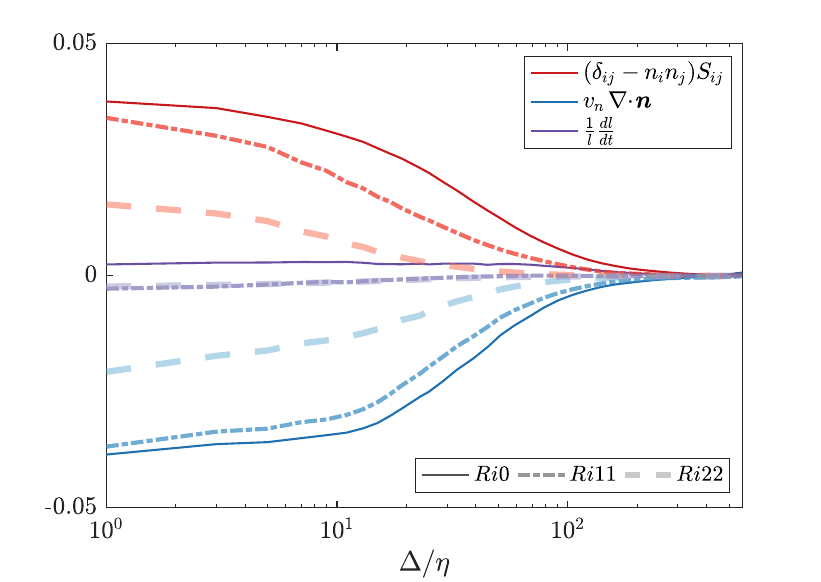}}
  \caption{Box and time averages of space-filtered stretching term (red), curvature/propagation term (blue) and the time evolution of a non-material infinitesimal element of area $\delta l$ (purple) for $30<t<120$.}
\label{fig:fig8}
\end{figure}

As observed in figure \ref{fig:fig2}, the positive and the negative patches of the terms in (\ref{equation:eq2}) appear to correlate with the orientation of the TNTI bulges. Since the TNTI surface has a fractal shape, in figure \ref{fig:fig8} we investigate the scale dependence of the terms in (\ref{equation:eq2}).
Following the procedure used in \citet{krug2017fractal}, we use a box filter to eliminate the effect of the spatial scales smaller than the size of the filter length. To find the position of the interface in the filtered field, we first convert the enstrophy field $\omega^{2}$ to a binary field $I$, where $I=1$ if $\omega^{2}>{\omega_{thr}}^2$  and $I=0$ if $\omega^{2}<{\omega_{thr}}^2$. We then filter $I$ according to $ \tilde{f}= \int f(\boldsymbol{x}-\boldsymbol{x^{\prime}}) $G$ (\boldsymbol{x^{\prime}}) d \boldsymbol{x^{\prime}} $, where $\tilde{f}$ is the filtered quantity and $G$ denotes the kernel of a square box filter of width ${\Delta}$. We then define the position of the filtered interface as the isocountour $I=0.5$. Contrary to filtering the enstrophy field directly, this procedure has the advantage that it preserves the mean position of the TNTI. As highlighted by \citet{krug2017fractal}, this is a necessary condition to keep the entrained flux across scales constant. Moreover, to compute the filtered terms in equation (\ref{equation:eq2}), we apply the same filter to the streamwise and the wall-normal velocity fields and evaluate the quantities in equation (\ref{equation:eq2}) at the location of the filtered interface.
 
In figure \ref{fig:fig8}, we display the time and space averages of the filtered quantities of equation (\ref{equation:eq2}) for different sizes of the filter length $\Delta$, here normalized with the Kolmogorov length scale $\eta$. All three-flow cases shown here display a similar behavior, with decaying magnitude of the stretching and the curvature/propagation terms with increasing filter size. Initially for $\Delta/\eta$ smaller than $\approx10$ the decay is very slow. Conversely, for $\Delta/\eta$ larger than $\approx10^{2}$ the magnitude of the stretching and the curvature/propagation terms is negligible, while for $\Delta/\eta$ between $\approx10$ and $\approx10^{2}$ a strong reduction of the two terms can be observed. We note that the limits of the latter region are consistent with the inner ($\approx 10\eta$) and the outer ($\approx 0.6-0.8h$) cutoffs of the scaling range of $A_{\eta }$ observed by \citet{krug2017fractal}.  Again, as seen in figure \ref{fig:fig5}, when $Ri$ increases, the terms on the rhs of equation (\ref{equation:eq2}) are smaller. The sum of the two terms of (\ref{equation:eq2}), the variation of the infinitesimal area, remains constant and close to 0 for all the filter sizes. That is to say, the two terms on the rhs of equation (\ref{equation:eq2}) not only balance out overall, but this balance holds also on a scale by scale basis. 

In conclusion, we observe that the difficulty on finding a time scaling for the stretching and the curvature/propagation term discussed in the context of figure \ref{fig:fig5} is in agreement with the results presented here, in that we showed how these terms are the result of multiscale process. They can therefore not be characterized by a single time scale.

\subsection{Impact of OECSs on the production/destruction of TNTI area}\label{subsec:OECSsTNTI}

A connection between local interface shape and nearby vortical structures was already established in \citet{neamtu2019lagrangian}. This is confirmed in figure \ref{fig:fig3} where we observed that part of the bulges of the TNTI are filled with OECSs. We now examine how the area change of the TNTI is related to the nearby coherent structures.

\begin{figure}
  \centerline{\includegraphics[width=1\linewidth]{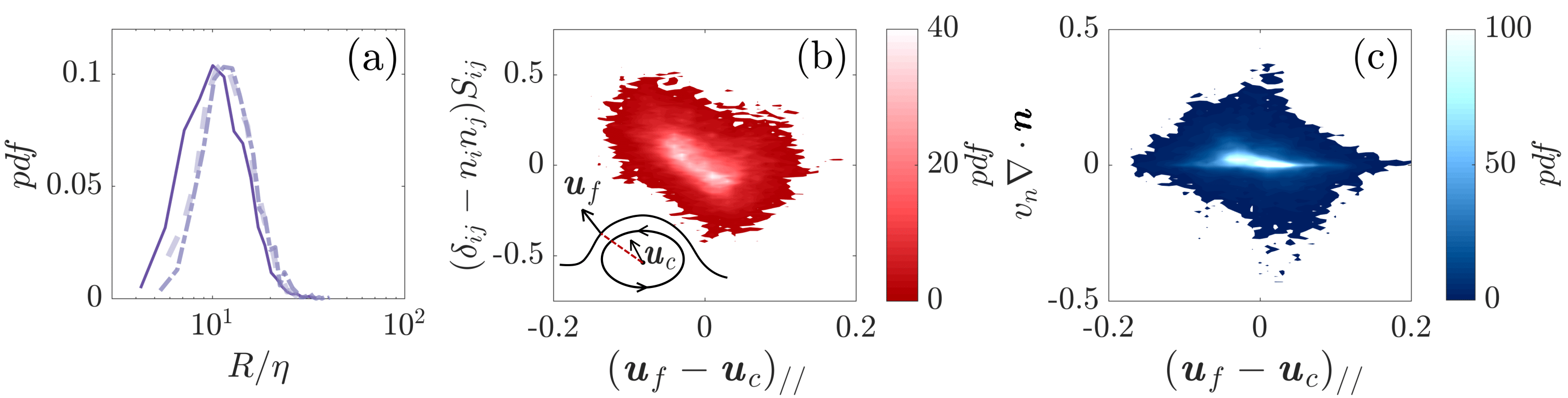}}
  \caption{PDF of the size (a) of the structures in the proximity of the TNTI. Increasing thickness and transparency of the curves corresponds to increasing $Ri$. Joint PDF of the stretching (b) and curvature/propagation terms (c) with respect to the relative velocity between the OECS and the TNTI for $Ri11$. The inset in (b) shows schematically where $\textbf{\textit{u}}_{f}$ and $\textbf{\textit{u}}_{c}$ are computed. }
\label{fig:fig9}
\end{figure}

As a first step, we focus only on the structures that shape the TNTI, that is, on the structures that are positioned `sufficiently' close to the TNTI. To this end, we compare the size $R$ of the structures with the minimum distance $\min(d)$ between the boundary of the structure and the TNTI. If the ratio $\min(d)/R$ is smaller than a threshold (here fixed at $1.5$), the structure is selected and discarded otherwise. The particular value of the threshold was chosen by observing that beyond this limit, the correlation between the structure position and the local interface evolution weakens considerably. In order to determine $R$, we fit an ellipse to the boundary of each structure and compute $R$ as the mean of the minor- and the major-axis of the ellipse. To give an impression of the typical size of the structures, we show the PDFs of the size of OECSs near the TNTI, normalized with the Kolmogorov length scale ${\eta}$ for all the flow cases in figure \ref{fig:fig9}(a). As can be seen, all three PDFs have a similar behavior, increasing rapidly from the minimum value of $R/{\eta} \approx 5$ to the position of the peak at about $R/{\eta} \approx 10$. The PDFs then decrease more slightly up to the highest values at about $R/{\eta} \approx 50$. 
Thus, the size of the OECSs covers a rather broad range of `active scales' over one order of magnitude, which may suggest an important role played by these structures in the multiscale aspects discussed in relation to figure \ref{fig:fig8}.

In order to understand the role of the OECSs on the evolution of the area of the TNTI, we examine whether the relative motion of the coherent structures with respect to the TNTI has an impact on the stretching, and the curvature/propagation term respectively. To this end, for each selected structure, we isolate a region of the TNTI in the neighborhood of the structure that lies within a distance from the center of the structure of about $5R$. This value was chosen in order to select most of the points of the bulge formed by the TNTI in the proximity of the OECS. We then compute the relative velocity $\textbf{\textit{u}}_{f}-\textbf{\textit{u}}_{c}$ between each of these points of the TNTI and the center of the OECS. Here $\textbf{\textit{u}}_{f}$ denotes the fluid velocity at the location of the TNTI and $\textbf{\textit{u}}_{c}$ the one at the location of the center of the OECS. We then project this velocity difference along the connection segment between each point of the TNTI and the center of the OECS to obtain $(\textbf{\textit{u}}_{f}-\textbf{\textit{u}}_{c})_{//}$. 
In figure \ref{fig:fig9}, we show the joint PDF of $(\textbf{\textit{u}}_{f}-\textbf{\textit{u}}_{c})_{//}$ against the strength of $(\delta_{ij}-n_{i}n_{j})S_{ij}$ (figure \ref{fig:fig9}b) respectively $v_{n} \nabla \cdot \textbf{\textit{n}}$ (figure \ref{fig:fig9}c). As can be gleaned from figure \ref{fig:fig9}(b), $(\textbf{\textit{u}}_{f}-\textbf{\textit{u}}_{c})_{//}$ and $(\delta_{ij}-n_{i}n_{j})S_{ij}$ are anti-correlated. That is, when $(\textbf{\textit{u}}_{f}-\textbf{\textit{u}}_{c})_{//}$ is negative, and thus the OECS approaches the TNTI, the stretching term is positive, whereas when $(\textbf{\textit{u}}_{f}-\textbf{\textit{u}}_{c})_{//}$ is positive the $(\delta_{ij}-n_{i}n_{j})S_{ij}$ contributes to the TNTI compression. This is not entirely unexpected given that the motion of the OECS towards (away from) the interface implies a compression (stretching) normal to the interface and hence, by incompressibility of water, stretching (expansion) in the tangential plane. A similar behavior is encountered for the curvature/propagation term and $(\textbf{\textit{u}}_{f}-\textbf{\textit{u}}_{c})_{//}$ (figure \ref{fig:fig9}c), although less pronounced than in the previous case. Also here, when the OECS approaches the TNTI the curvature/propagation term produces area, whereas when the OECS moves away from the interface, the TNTI area is decreased. Both the joint PDFs shown in figure \ref{fig:fig9} indicate that the motion of the OECS near the TNTI have an influence on the time evolution of the area of the TNTI. 

\subsubsection{Self similarity of the flow fields around OECSs near the TNTI}\label{subsubsec:SelfSim}

In the following, we demonstrate that the flow fields around the OECSs near the TNTI are self similar with respect to the size of the structures by means of a conditional analysis. In particular, we re-sample the velocity, the enstrophy and the IVD fields onto a uniform grid, frame referenced at the center of the OECSs, and normalized in $x$ and $z$ directions with the average size of the structures $R$. The rationale here is to have a common frame of reference for all the OECSs and to compare flow fields around OECSs of the same normalized size. By taking the average of the IVD fields around the OECSs, we extract a mean representative OECS, that is, we apply the extraction algorithm to the conditional average of the IVD field. To compute the conditional average of the TNTI position, we apply the same change of the coordinate system to the TNTI position near each selected OECS. By defining a curvilinear coordinate $s/R$ along the TNTI which has its origin at the same $x$ coordinate of the highest point on the boundary of the OECS, we compute $\tilde x_{i}(s)$ and $\tilde z_{i}(s)$, the normalized coordinates of the TNTI position near the $i$-th OECS. In this way, the mean position of the TNTI is computed by taking the average of  $\tilde x_{i}(s)$ and $\tilde z_{i}(s)$ over $i$ conditioning with respect to $s/R$. The same procedure is used also for the terms in (\ref{equation:eq2}), as well as for the entrainment velocity $v_{n}$.

\begin{figure}
  \centerline{\includegraphics[width=1\linewidth]{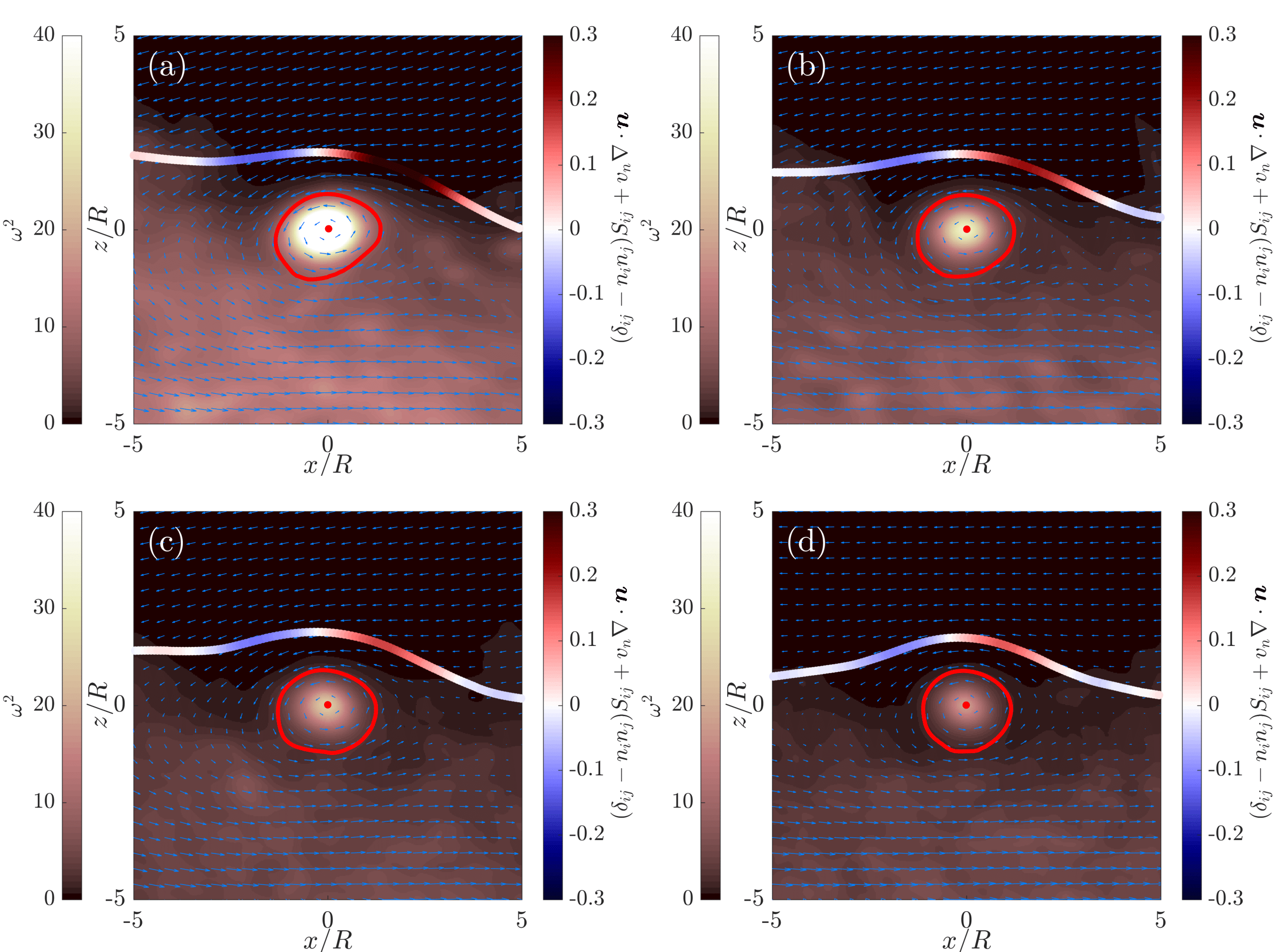}}
  \caption{Mean OECS and conditional average position of the TNTI for $Ri11$. The boundary of the mean OECS is constituted by the red closed-curve. The open curve is the conditional average position of the TNTI, color-coded with the time evolution of a non-material infinitesimal element of area $\delta l$. The direction and the size of the vectors represent the conditional average velocity field relative to the velocity at the center of the OECSs, whereas in the background the conditional average of the enstrophy field is shown. The conditional analysis is limited to structures near the TNTI with respectively $R/{\eta}<10.5$ (a), $10.5<R/{\eta}<12.75$ (b), $12.75<R/{\eta}<15$ (c) and $R/{\eta}>15$ (d).}
\label{fig:fig10}
\end{figure}

The normalization used here for the spatial coordinates presupposes self similarity of the velocity and the enstrophy fields in the proximity of the OECSs, as well as of the TNTI position and of the terms in (\ref{equation:eq2}) with respect to the size of the structures $R$. In order to demonstrate this self similarity, we subdivide the selected structures near the TNTI (of the $Ri11$ flow case) in four different groups with similar cardinality based on their size (with the limits indicated in the caption of figure \ref{fig:fig10}) and we conduct a conditional analysis using the spatial normalization indicated above.
 
In figure \ref{fig:fig10}, we present the results of the conditional analysis. Here, the mean position of the TNTI is color-coded with the conditional average of the time evolution of the surface area of the TNTI. Moreover, in the background the conditional average of the enstrophy field is shown, while the arrows indicate the average velocity field relative to the velocity at the center of the OECS. Focusing on figure \ref{fig:fig10}(a), the first observation that emerges is that the conditional average position of the TNTI is shaped by the nearby OECSs. Also, the mean flow around the OECS in the frame of reference co-moving with the center of the structures indicates a rotational fluid motion. The average enstrophy field is particularly high at the center of the mean OECS, decreasing radially towards the boundaries of the OECS. Furthermore, the conditional average of the infinitesimal area variation (shown in colors at the mean position of the TNTI) displays a clear pattern: positive at the leading edge and negative at the trailing edge. As can be seen in figure \ref{fig:fig10}(b-d), all the groups show identical patterns and there is a striking similarity in all the aspects discussed for figure \ref{fig:fig10}(a). However, as the size of the structures increases (from figure \ref{fig:fig10}a to figure \ref{fig:fig10}d), both the intensities of the enstrophy field inside the boundaries of the OECS and of the flow field around the OECS decrease. This means that the smaller the structures are, the faster is their rotational motion. Moreover, also the typical pattern shown by the conditional average of the time evolution of the TNTI appears to weaken, when the size of the structures is larger. In conclusion, the results shown in figure \ref{fig:fig10} demonstrate clearly that the vortical structures near the TNTI exhibit a scale-invariant behavior. 
\begin{figure}
  \centerline{\includegraphics[width=1\linewidth]{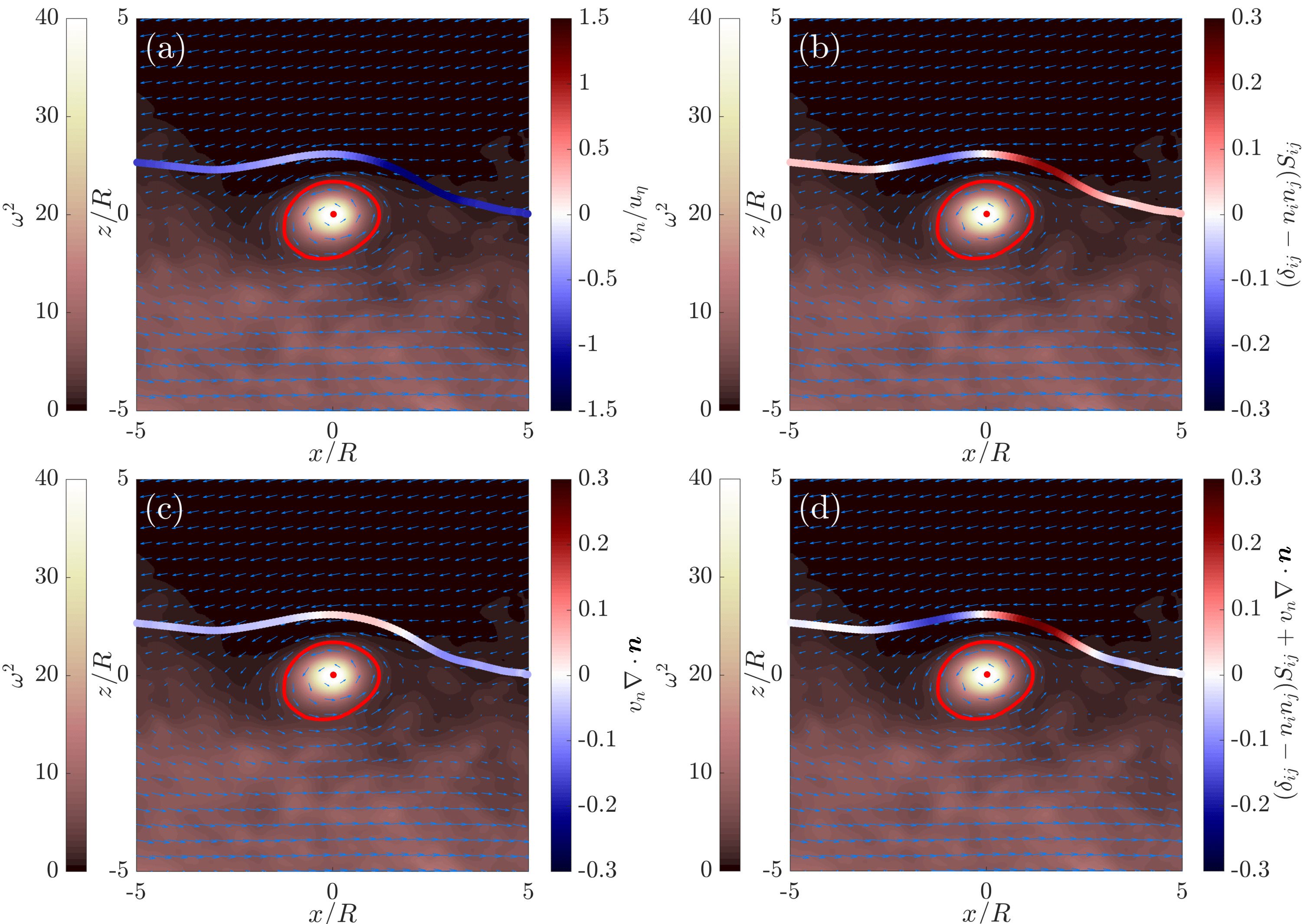}}
   \caption{Same conditional average shown in figure \ref{fig:fig10} for $Ri11$. In this case, the conditional average position of the TNTI is color-coded with the average entrainment velocity (a), stretching (b) and curvature/propagation terms (c) and the time evolution of a non-material infinitesimal element of area $\delta l$.}
\label{fig:fig11}
\end{figure}

\subsubsection{Conditional analysis of the time evolution of the TNTI}\label{subsubsec:CondAnalysisTNTI}

The effect of the OECSs on the terms of (\ref{equation:eq2}) is investigated in the following through the conditional analysis presented in section \S\ref{subsubsec:SelfSim}. In figure \ref{fig:fig11}, we present the results of the conditional analysis for $Ri11$ in which all the selected structures near the TNTI are considered. 
\begin{figure}
  \centerline{\includegraphics[width=0.65\linewidth]{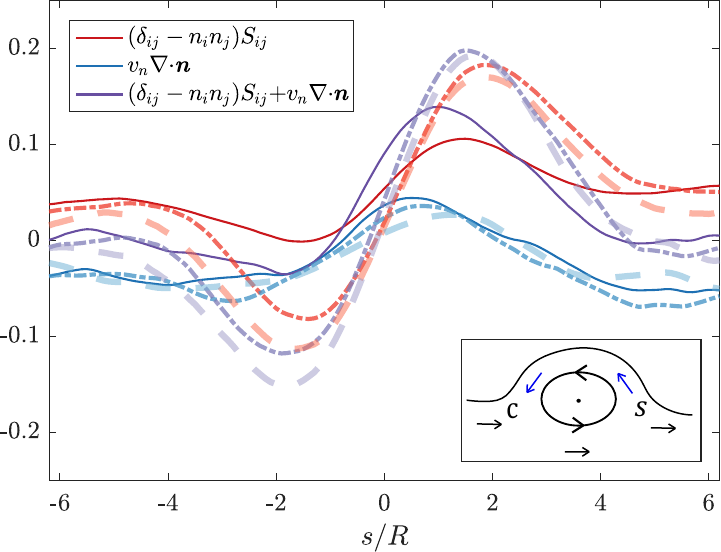}}
  \caption{Conditional average of the stretching (blue) and the curvature/propagation term (red) and of their sum (purple) in the proximity of the OECS against the curvilinear abscissa $s/R$ for $Ri0$ (continuous line), $Ri11$ (dash-dotted line) and $Ri22$ (dashed line). In the inset, a schematic of the conditional average flow direction around the OECS in the proximity of the interface is shown. The arrows indicate the flow direction between the boundary of the OECS and the TNTI (blue) and in the turbulent region below the OECS (black). `$C$' denotes the compression, whereas `$S$' denotes the stretching.}
\label{fig:fig12}
\end{figure}
In particular, in figure \ref{fig:fig11}(a), the mean TNTI is color-coded with the conditional average of the entrainment velocity conditioned with respect to the position on the TNTI. As can be seen, $v_{n}$ varies along the TNTI and in particular, it is negative at the leading edge, where it reaches values as high as $v_{n}/u_{\eta}=-1$, whereas it decreases (in magnitude) at the trailing edge. This result is consistent with the findings from the experimental work by \citet{neamtu2019lagrangian}, where a very similar pattern for $v_{n}$ was observed. In figure \ref{fig:fig11}(b), we display the conditional average of the stretching term of equation (\ref{equation:eq2}). Here, the mean stretching term has a clear pattern along the TNTI. Indeed, it has positive values at the leading edge respectively negative values at the trailing edge. On the tails, far from the center of the structure, it shows smaller and positive values recovering the unconditioned average value. The surface of the TNTI shown in figure \ref{fig:fig11}(c) is color-coded the conditional average of the curvature/propagation term of equation (\ref{equation:eq2}). The mean curvature/propagation term exhibits a similar behavior as the stretching term, being slightly positive at the leading edge and negative at the trailing edge. The sum of the two terms is plotted in figure \ref{fig:fig11}(d). The average infinitesimal area growth rate shown here is positive at the leading edge and negative at the trailing edge. That is, the interface area is produced at the leading edge and destroyed at the trailing edge, whereas far from the structure is negligible reaching the unconditioned average value. 

In the following, a comparison of the conditional averages of terms in (\ref{equation:eq2}) among the different flow cases is presented. For an easier quantitative analysis, in figure \ref{fig:fig12} the conditional averages are shown against the curvilinear abscissa $s/R$. As can be noticed, all three terms of equation (\ref{equation:eq2}) have similar trends independently of the flow cases. In particular, moving from $s/R=-6$ in the direction of increasing $s/R$, the stretching term decreases, reaching a minimum at about $s/R\approx -2$.  Continuing in the in the direction of increasing $s/R$, $(\delta_{ij}-n_{i}n_{j})S_{ij}$ starts to increase and after an inflection point at $s/R=0$, it reaches a maximum at $s/R\approx +2$ and then it decreases again. The behavior just described is more intense for the stratified cases, where the minimum of $(\delta_{ij}-n_{i}n_{j})S_{ij}$ at $s/R\approx -2$ is negative. For $Ri0$, the $(\delta_{ij}-n_{i}n_{j})S_{ij}$ has the same behavior but it never reaches negative values. Interestingly, the inflection point of $(\delta_{ij}-n_{i}n_{j})S_{ij}$ happens exactly at $s/R=0$. A possible mechanism that generates the tangential stretching/compression at the leading/trailing edge of the TNTI surface near the vortical structures is schematically depicted in the inset of figure \ref{fig:fig12}. Here the arrows indicate the conditional average flow direction of the fluid motion between boundary of the OECS and the TNTI (blue) and in the turbulent region below the OECS (black). As can be seen, at the leading edge the flow tends to stretch fluid elements, when projected on the tangent to the TNTI surface. Conversely, at the trailing edge, fluid elements are compressed along the direction tangent to the TNTI. In the mechanism just described, the faster the fluid layers beneath the structures, the higher the difference between the relative maximum and minimum of the stretching term. As $Ri$ increases, faster and more horizontal layers of fluid are known to form in the turbulent region. These layers might be responsible for the observed difference among the flow cases. We also note that in the conditional average stretching shown in figure \ref{fig:fig12}, the positive contribution of the larger scales is superimposed to the mechanism just described. This is why, for $Ri0$ the relative minimum of the stretching term never reaches negative values. Finally, it is worth noting that the formation of the positive-negative double-peak seen for $Ri22$ is compatible with the more symmetric stretching observed in figure \ref{fig:fig6}.

Starting from $s/R=-6$ where it is negative and moving into direction of positive $s/R$, the conditional average of the curvature/propagation term increases, reaching a positive maximum at  $s/R\approx 1$, and decreases, turning again negative at about $s/R=2$. To explain this behavior, we notice that on both sides of a bulge the curvature is positive (valleys), whereas the entrainment velocity is negative (ambient fluid is entrained). Thus their product is negative and it contributes to the destruction of the area of the TNTI. This is why, negative values of the curvature/propagation term are observed at both sides of the OECSs. Moving from downstream to the leading edge, the curvature of the interface changes the sign (at about $s/R \approx 2$) becoming negative (concave shape as seen form the turbulent side). The entrainment velocity is still negative here, and thus the curvature/propagation term becomes positive. A weak effect of the stratification can be noticed also here where a higher positive value for the maximum of $v_{n} \nabla \cdot \textbf{\textit{n}}$ can be observed for the unstratified case.

The sum of the two terms, the time variation of the infinitesimal area, shows a profile that is similar to the stretching term. Indeed in the proximity of the structure, $(\delta_{ij}-n_{i}n_{j})S_{ij}$ dominates the time evolution of the interface area, dominating over $v_{n} \nabla \cdot \textbf{\textit{n}}$. Thus, the time variation of the infinitesimal area is positive at the leading edge, it has an inflection point at about $s/R \approx 0$ and it becomes negative at the trailing edge. That is, the area of the TNTI is produced at the leading edges and destroyed at the trailing edges. This behavior is clearly consistent with the trends observed qualitatively in figure \ref{fig:fig2}.

\section{Concluding remarks}\label{sec:discussion}

In this paper, we investigated the production/destruction process of the TNTI area of gravity current flows. We showed that a curvature/propagation term, which is particularly active in the valleys of the TNTI (figure \ref{fig:fig2}), contributes on average to the destruction of the TNTI area (figure \ref{fig:fig5}), while this is counterbalanced by a flow stretching which on average produces interface area (figure \ref{fig:fig5}). In particular, the stretching drives the time evolution of the TNTI area for weak to moderate events, whereas the curvature/propagation term dominates the extreme events (figure \ref{fig:fig6}). Very similar results have been found by \citet{wang2017direct} in the case of a premixed jet flame. In their work, the authors report that the tangential strain produces interface in regions with low curvature, whereas the curvature/propagation term destroys flame surface area in high curvature regions.

The multiscale aspects of the production/destruction of the TNTI area were investigated by filtering the scales smaller than a filter length ${\Delta}$ (figure \ref{fig:fig8}). We showed that the stretching and the curvature/propagation terms balance each other at all scales and that the magnitude of the two terms decreases with increasing ${\Delta}$. Recently, \citet{mistry2016entrainment} showed that the mass-flux rate of entrained fluid across the TNTI is constant across all the scales ${\Delta}$, as envisaged by \citet{meneveau1990interface}. That is, the reduction of $A_{\eta}$ with increasing ${\Delta}$ is balanced by the enhancement of $v_{n}$ at the larger scales. In our case, $v_{n} \nabla \cdot \textbf{\textit{n}}$ reduces with increasing filter length, which means that the increment of $v_{n}$ is smaller than the rapid decrease of curvature with increasing ${\Delta}$. 

We showed that increasing stratification reduces the averages of both the stretching and the curvature/propagation terms (figure \ref{fig:fig5}), while maintaining the same trends for their time variation (figure \ref{fig:fig5}) and their filtered values (figure \ref{fig:fig8}). In particular, we showed that the reduction of the stretching term is largely attributable to changes in $\delta_{ij}-n_{i}n_{j}$ tensor (figure \ref{fig:fig7}) and we associated this with the tendency of the interface to flatten with increasing stratification. Indeed, as shown by \citet{krug2017fractal}, for the same flows investigated here, the convolutions of the TNTI scale with the shear length scale $l_{sk}=e^{1/2}/\overline{S}$ in the vertical direction respectively with $h$ in the streamwise direction. While the range of the length scales of the TNTI convolutions impacted by the stable stratification remains the same, it is the growing anisotropy implied by different scalings that modify the geometry of the interface with increasing $Ri$. Moreover, this change of the TNTI geometry is also compatible the observation that the positive tail of the interface curvature reduces with increasing $Ri$, which together with the entrainment velocity reduction explains the reduction of the curvature/propagation term. 

The local effect of the coherent structures on the TNTI area production/destruction process was investigated through a conditional analysis similar to \citet{neamtu2019lagrangian}. The conditional analysis showed that both stretching and the curvature/propagation terms produce TNTI area at the leading edge and to destroy it at the trailing edge of the TNTI in proximity of vortical structures. In particular, we inferred that the behavior of the stretching term is related to the mean rotational motion induced by the vortical structures (figure \ref{fig:fig12}). A similar mechanism for the tangential stretch of a premixed flame was described in \citet{sinibaldi2003propagation}, where the authors observed that toroidal vortices near the flame boundaries induce a rotational motion that stretches the boundaries on one side of the vortices and compresses it on the other. 

Using the conditional analysis, we demonstrated the existence of a scale invariant behavior of the vortices near the TNTI  (figure \ref{fig:fig10}). A self-similarity of the shapes of the TNTI has been observed a long time ago \citep{sreenivasan1989mixing}, however a similar observation for the coherent structures near the TNTI is missing up to date in the literature. Moreover, we showed that also the conditional average of the time evolution of the TNTI is self-similar with respect to the size of the OECS. In particular, the time evolution of the TNTI area exhibits lower values for increasing size of the OECS. In figure \ref{fig:fig12}, we displayed that near the vortical structures, the time evolution of the interface is mostly dominated by the stretching term. This together with the observation that the rotational motion of the OECS is slower for increasing size of the OECSs (figure \ref{fig:fig10}) lends support to the interpretation that the stretching mechanism of the interface is related to the rotation of the OECSs.

In conclusion, to our knowledge, the detailed analysis presented here constitutes a first tentative to describe the evolution of the surface area of the TNTI in the case of turbulent stably stratified shear flows. This work may motivate further studies into the production/destruction mechanisms of the TNTI of other types of turbulent shear flows, such as turbulent jets and wakes and turbulent boundary layers. In particular, it may be of interest to understand whether the vortical structures near the TNTI of these flows have a similar impact on the time evolution of the TNTI and universally apply to all shear flows with an interface. A further issue that may deserve further study is how stratification acts to suppress stretching at large inertial scales.
\\
\\
\textbf{Acknowledgements}

We are grateful for financial support from DFG priority program SPP 1881 under grant number HA 7497/1-1.

J.-P.M. and M.v.R. were supported by the EPSRC project Multi-scale Dynamics at the Turbulent/Non-turbulent Interface of Jets and Plumes (grant number EP/R043175/1) and the UK Turbulence consortium (grant number EP/R029326/1).
\\
\\
\textbf{Declaration of Interests}

The authors report no conflict of interest.
%



\bibliographystyle{jfm}

\end{document}